\documentclass[12pt]{article}
\begin{document}
\begin{center}
\large\bf{The Bargmann-Wigner Equations in\\
Spherical Space}
\end{center}
\vspace{2cm}
\begin{center}
D.G.C. McKeon$^{(a,b)}$\\
T.N. Sherry$^{(b)}$\vspace{1cm}
\begin{enumerate}
\item[(a)] Department of Applied Mathematics\\
University of Western Ontario\\
London  CANADA\\
N6A 5B7
\item[(b)] Department of Mathematical Physics\\
National University of Ireland Galway\\
Galway  Ireland
\end{enumerate}
\end{center}
\vspace{1cm}
email: dgmckeo2@uwo.ca $\;\;\;\;$
tom.sherry@nuigalway.ie\\
Tel: (519)679-2111, ext. 88789\\
Fax: (519)661-3523 \vspace{4cm}
\begin{flushright}
\hfill PACS 11:30-j
\end{flushright}
\eject
\section{Abstract}

The Bargmann-Wigner formalism is adapted to spherical surfaces embedded in three to eleven dimensions. This is demonstrated to generate wave equations in spherical space for a variety of antisymmetric tensor fields. Some of these equations are gauge invariant for particular values of parameters characterizing them. For spheres embedded in three, four and five dimensions, this gauge invariance can be generalized so as to become non-Abelian. This non-Abelian gauge invariance is shown to be a property of second order models for two index antisymmetric tensor fields in any number of dimensions. The $O(3)$ model is quantized and the two point function shown to vanish at one loop order.

\section{Introduction}

The Bargmann-Wigner (BW) equations [1,2] are a means of generating wave equations for higher spin fields. For spin $s$, a totaly symmetric $2s$ component spinor $\psi_{\alpha_{1}\alpha_{2} \ldots \alpha_{2s}}$ is taken to satisfy the $2s$ equations
$$(i\gamma \cdot \partial + m)_{\alpha , \beta}\, \psi_{\beta\alpha{_2}\cdots\alpha_{2s}} = 0 \ldots 
(i\gamma \cdot \partial + m)_{\alpha_{2s} \beta} \psi_{\alpha_1\alpha{_2}\cdots\beta} = 0.\eqno(1)$$
(We work in Euclidean space with $\gamma$-matrix conventions given in the appendix.)

Before adapting this procedure to spherical space, we will demonstrate how eq. (1) can be used in the case $s = 1$ to generate the Maxwell equations. By eqs. (A.15) and (A.16), we see that
$$\psi_{\alpha\beta} (x) = m A_a (x) \left(\gamma_a\gamma_5C_5\right)_{\alpha\beta} - {^*}F_{ab}(x) \left(\Sigma_{ab} C_5\right)_{\alpha\beta}\;.\eqno(2)$$

Eqs. (A7-A10) show that together eqs. (1) and (2) lead to
$$\left[im\, \partial_a A_b \left(\delta_{ab} + 2i \Sigma_{ab}\right) + m^2 \left(A_a\gamma_a\right)\right]
+ \left[\frac{1}{2} \partial_a F_{bc} \left(\delta_{ab} \gamma_c - \delta_{ac} \gamma_b - \epsilon_{abcd}\gamma_d\gamma_5\right) + m F_{ab} \Sigma_{ab}\right] = 0.\eqno(3)$$
We have set ${^*}F_{ab} = \frac{1}{2} \epsilon_{abcd} F_{cd}$ so that ${^*}\left({^*}F_{ab}\right) = F_{ab}$ and ${^*}\Sigma_{ab} = -\Sigma_{ab}\gamma_5$. By eq. (3) we recover the standard Proca equations
$$\partial_a F_{ab} + m^2 A_b = 0\eqno(4)$$
$$F_{ab} = \partial_a A_b - \partial_bA_a\eqno(5)$$
$$\partial \cdot A = 0\eqno(6)$$
$$\partial_a {^*}F_{ab} = 0.\eqno(7)$$
For the particular value $m^2 = 0$, eqs. (4) and (5) become the free field Maxwell equations; they possess the gauge invariance
$$A_a \rightarrow A_a + \partial_a\theta .\eqno(8)$$
Eq. (6) is not gauge invariant; it corresponds to a gauge fixing condition.

The form of Dirac equation on a spherical surface is [3-4]
$$\left(\Sigma_{ab} L_{ab} + \xi\right)_{\alpha\beta} \psi_\beta = 0.\eqno(9)$$
This suggests that we consider the following generalization of eq. (1) for a wave function $\psi_{{\alpha{_1}}\ldots \alpha_{2{_s}}}$ that is symmetric in all its subscripts,
$$\left(\Sigma_{ab} L_{ab} + \xi\right)_{\alpha ,\beta} \psi_{\beta\alpha_2 \ldots \alpha_{2s}}  = 0\eqno(10)$$
$$\!\!\!\!\!\!\!\!\!\!\!\!\!\!\!\!\!\!\!\!\!\!\!\dots\nonumber$$
$$\left(\Sigma_{ab} L_{ab} + \xi\right)_{\alpha_{2s}\beta} \psi_{\alpha_1\alpha_2 \ldots \beta}  = 0.\nonumber$$
We show below that for the case $s = 1$, ensuring symmetry in $(\alpha , \beta)$ for the bispinor wave function $\psi_{\alpha\beta}$ results in a set of wave equations for wave functions that are antisymmetric in spatial indices. Subsequently this is shown explicitly in $n$ dimensional spherical spaces $S_n$ for $n = 2, 3, 4, 5, 6, 7, 8, 10$.  In a number of these spaces, the equations possess an Abelian gauge invariance for particular values of the parameter $\xi$ in eq. (10). This gauge invariance can be generalized so as to become non-Abelian when the wave function is antisymmetric in two spatial indices. Actions that possess this non-Abelian gauge invariance can be formulated.  Finally a model on $S_2$ is quantized and it is shown that at one loop order in this model the two point function vanishes. Quantization of an $S_3$ model is briefly discussed.

Wave functions that are antisymmetric in their spatial indices naturally arise in string theories [5]. This is what motivated us to consider their properties on spherical surfaces $S_n$. Using isometry generators on $S_n$ defined in terms of $n$ coordinates of the $(n + 1)$ dimensional embedding space makes it possible to formulate gauge invariant versions of these models.

Other discussions of gauge fields on spaces of constant curvature have appeared in the literature [6].

\section{Bargmann-Wigner Equations on $S_n$}
\subsection{$S_2$}

The equation for a BW wave function $\psi_{\alpha\beta}$ on $S_2$ can be reduced from the form of eq. (10) to simply
$$\left(\vec{\tau} \cdot \vec{L} + \xi\right)_{\alpha\beta} \psi_{\beta\gamma} = 0\eqno(11)$$
upon setting
$$L_i = -i\epsilon_{ijk} \eta_j \frac{\partial}{\partial\eta_k} = -i\left(\vec{\eta} \times \vec{\nabla}\right)_i\;.\eqno(12)$$
(The Dirac equation $\left(\vec{\tau} \cdot \vec{L} + \xi\right)_{\alpha\beta}\psi_{\beta}$ arose on $S_2$ in ref [7].)  

If
$L_{ij} = \epsilon_{ijk} L_k$, $\Sigma_{ij} = \frac{1}{2} \epsilon_{ijk} \tau_k$, then eq. (11) can be written as
$$\left(\Sigma_{ij} L_{ij} + \xi\right)_{\alpha\beta} \psi_{\beta\gamma} = 0\eqno(13)$$
which is of the form of eq. (10).

By eq. (A.12), we see that
$$\psi_{\alpha\beta} (x) = \phi_i (x) \left(\tau_i C_3\right)_{\alpha\beta} ;\eqno(14)$$
together eqs. (11) and (14) lead to
$$i\epsilon_{ijk} L_j \phi_k + \xi \phi_i = 0\eqno(15)$$
$$L_i\phi_i = 0.\eqno(16)$$
Provided
$$\xi = 1, \eqno(17)$$
eq. (15) is invariant under the transformation
$$\phi_i \rightarrow \phi_i + i L_i\beta .\eqno(18)$$

If $\phi_{ij} = \epsilon_{ijk} \phi_k$, then eqs. (15), (16) and (18) become
$$i\left(L_{pi} \phi_{pj} - L_{pj} \phi_{pi}\right) + \xi\phi_{ij} = 0\eqno(19)$$
$$L_{ij}\phi_{ij} = 0\eqno(20)$$
and
$$\phi_{ij} \rightarrow \phi_{ij} + i L_{ij} \beta\eqno(21)$$
respectively.

An action which gives rise to the equation of motion of eq. (15) (with eq. (17) holding) is
$$S = \int d\Omega \vec{\phi} \cdot \left(i\vec{L} \times \vec{\phi} + \vec{\phi}\right).\eqno(22)$$
A non-Abelian generalization of eq. (22) is
$$S = \int d\Omega \left[ \vec{\phi}^a \left(i \vec{L} \times \vec{\phi}^a + \vec{\phi}^a\right)  + \frac{1}{3} \epsilon^{abc} \vec{\phi}^a \cdot \vec{\phi}^b \times \vec{\phi}^c\right].\eqno(23)$$
This is invariant under the infinitesmal $SO(3)$ transformation
$$\phi_i^a \rightarrow \phi_i^a + iL_i\beta^a + \epsilon^{abc}\phi_i^b\beta^c.\eqno(24)$$
It is also possible to define a ``field strength'' 
$$f_i^a = i \epsilon_{ijk} L_j\phi_k^a + \phi_i^a + \frac{1}{2} \epsilon^{abc} \epsilon_{ijk} \phi_j^b \phi_k^c ;\eqno(25)$$
under the gauge transformation of eq. (24), we find that
$$f_i^a \rightarrow f_i^a + \epsilon^{abc} f_i^b \beta^c .\eqno(26)$$
Consequently, the action
$$S = k\int d\Omega  f_i^a f_i^a\eqno(27)$$
possesses non-Abelian gauge invariance.

\subsection{$S_3$}

The form of a symmetric wave function $\psi_{\alpha\beta}$ in a four dimensional embedding space is, by eqs. (A.15) and (A.16)
$$\psi_{\alpha\beta} = V_i\left(\gamma_i\gamma_5C_5\right)_{\alpha\beta} - {^*}\phi_{ij}\left(\Sigma_{ij} C_5\right)_{\alpha\beta}
\left({^*}\phi_{ij} \equiv \frac{1}{2} \epsilon_{ijk\ell}\phi_{k\ell}\right).\eqno(28)$$
The BW eq. (10) now leads to
$$i\left(L_{ki} \phi_{kj} - L_{kj} \phi_{ki}\right) + \xi\phi_{ij} = 0\eqno(29)$$
$$\epsilon_{ijk\ell} L_{ij} \phi_{k\ell} = 0 \eqno(30)$$
$$L_{ij} \phi_{ij} = 0\eqno(31)$$
$$i L_{ki} V_k + \xi V_i = 0\eqno(32)$$
$$\epsilon_{ijk\ell} L_{jk} V_\ell = 0.\eqno(33)$$

The fields $\phi_{ij}$ and $V_i$ decouple on $S_3$, unlike the fields $A_a$ and $F_{ab}$ in eqs. (4-7). Eqs. (29) and (30) are invariant under the gauge transformation
$$\phi_{ij} \rightarrow \phi_{ij} + iL_{ij}\theta \eqno(34)$$
provided
$$\xi = 2.\eqno(35)$$
An action consistent with the gauge invariance of eq. (24) is
$$S = \int d\Omega \left[ \phi_{ij} \left( iL_{ki} \phi_{kj} + \phi_{ij}\right) + \lambda \epsilon_{ijk\ell} L_{ij} \phi_{k\ell}\right].\eqno(36)$$
The equation of motion associated with the field $\lambda$ is eq. (30); however the equation of motion associated with $\phi_{ij}$ now becomes
$$i\left(L_{ki} \phi_{kj} - L_{kj} \phi_{ki}\right) + 2\phi_{ij} = \epsilon_{ijk\ell} L_{k\ell} \lambda \eqno(37)$$
in place of eq. (19). Only if $\lambda = 0$ and $\xi = 2$ do eqs. (29) and (37) coincide.

The gauge invariance of eq. (34) can be generalized to an infinitesimal $SO(3)$ non-abelian gauge invariance
$$\phi_{ij}^a \rightarrow \phi_{ij}^a + iL_{ij} \theta^a + \epsilon^{abc} \phi_{ij}^b \theta^c ;\eqno(38)$$
$$S = \int d\Omega \left[ \phi_{ij}^a \left(iL_{ki} \phi_{kj}^a + \phi_{ij}^a\right) + \frac{1}{3} \epsilon^{abc} \phi_{ij}^a \phi_{jk}^b \phi_{ki}^c + \lambda^a \epsilon_{ijk\ell} L_{ij} \phi_{k\ell}^a\right] \eqno(39)$$
is invariant under the transformation of eq. (38) provided
$$\lambda^a \rightarrow \lambda^a + \epsilon^{abc}\lambda^b \theta^c .\eqno(40)$$
As in eq. (27), the ``Lagrange multiplies"' field $\lambda^a$ enters into the equation of motion for the field $\phi_{ij}^a$.

In $D$ dimensions (ie, on the sphere $S_{D-1}$) the ``field strength''
$$f_{ij}^a \equiv i\left( L_{ki} \phi_{kj}^a - L_{kj}\phi_{ki}^a\right) + ( D-2) \phi_{ij}^a\eqno(41)$$
$$+ \frac{1}{2} \epsilon^{abc} \left(\phi_{ki}^b \phi_{kj}^c - \phi_{kj}^b \phi_{ki}^c\right)\nonumber$$
is covariant under the gauge transformation of eq. (38),
$$f_{ij}^a \rightarrow f_{ij}^a + \epsilon^{abc} f_{ij}^b \theta^c ,\eqno(42)$$
as by eq. (A5)
$$\left[ L_{ki}, L_{kj}\right] = i(D-2) L_{ij} .\eqno(43)$$
Consequently, the action
$$S = k\,\int d\Omega  f_{ij}^a f_{ij}^a \eqno(44)$$
possesses the gauge invariance of eq. (38). The last term in eq. (39) could also be incorporated into eq. (44).

On $S_3$, if the equation of motion that follows from this gauge invariant action is satisfied
$$i\left(L_{ki} f_{kj}^a - L_{kj}f_{ki}^a\right) + 2f_{ij}^a + \epsilon_{abc} \left(\phi_{ki}^b f_{kj}^c - \phi_{kj}^b f_{ki}^c\right) = 0\eqno(45)$$
it follows that
$$i\left(L_{ki} {^*}f_{kj}^a - L_{kj}{^*}f_{ki}^a\right) + {^*}f_{ij}^a + \epsilon_{abc} \left(\phi_{ki}^b {^*}f_{kj}^c - \phi_{kj}^b {^*}f_{ki}^c\right) = 0.\eqno(46)$$
This is a consequence of the identities
$${^*}({^*}A_{ij}) = A_{ij}\eqno(47)$$
and
$${^*}\left(A_{ki} {^*}B_{kj} - A_{kj} {^*}B_{ki}\right) = A_{ki} B_{kj} - A_{kj}B_{ki}\eqno(48)$$
provided $A_{ij} = -A_{ji}$ and $B_{ij} = -B_{ji}$.
\subsection{$S_4$}

By eq. (A17), the only contribution to the BW wave function on $S_4$ is
$$\psi_{\alpha\beta} = \phi_{ij} \left(\Sigma_{ij} C_5\right)_{\alpha\beta}\;.\eqno(49)$$
Upon using eqs. (A4) and (A6), eqs. (10) and (49) together lead to 
$$i\left(L_{ki} \phi_{kj} - L_{kj} \phi_{ki}\right) + \xi \phi_{ij} = 0\eqno(50)$$
$$\epsilon_{ijk\ell m} L_{jk} \phi_{\ell m} = 0\eqno(51)$$
$$L_{ij} \phi_{ij} = 0\;.\eqno(52)$$
Eqs. (50) and (51) are invariant under the gauge transformation
$$\phi_{ij} \rightarrow \phi_{ij} + i L_{ij} \theta\eqno(53)$$
if
$$\xi = 3 .\eqno(54)$$
This gauge symmetry is present in the action
$$S = \int d\Omega \left[\phi_{ij}\left(i L_{ki} \phi_{kj} + \frac{3}{2} \phi_{ij}\right) + X_i \epsilon_{jk\ell m} L_{jk} \phi_{\ell m}\right].\eqno(55)$$

In addition, there is a vector gauge invariance which is identical to that of vector fields in ref. [4]
$$X_i \rightarrow X_i + i \eta_j L_{ij}\omega .\eqno(56)$$
Again, the equation of motion for $\phi_{ij}$ generates not just eq. (50) but also an extra term dependent on the Lagrange
multiplier field $X_i$.

An immediate generalization so that the infinitesimal $SU(2)$ non-Abelian gauge invariance of eq. (38) is present on $S_4$ is
$$S = \int d\Omega \left[ \phi_{ij}^a \left(i L_{ki} \phi_{kj}^a + \frac{3}{2} \phi_{ij}^a\right)\right.\nonumber$$
$$\left. +\frac{1}{3} \epsilon^{abc} \phi_{ij}^a \phi_{jk}^b\phi_{ki}^c + X_i^a \epsilon_{ijk\ell m} L_{jk} \phi_{\ell m}^a \right] \eqno(57)$$
provided that now
$$X_i^a \rightarrow X_i^a + \epsilon^{abc} X_i^b \theta^c.\eqno(58)$$
Again, eqs. (41) and (44) can be used with $D = 5$ to generate an action with non-Abelian gauge invariance on $S_4$.
\subsection{$S_5$}

By eqs. (A.21 - A.24), the form of a BW wave function on $S_5$ is
$$\psi_{\alpha\beta} = \sigma (C_7)_{\alpha\beta} + \phi_{ij} \left(g_{ij}g_7C_7\right)_{\alpha\beta} + \tau_{ijk}\left(g_{ijk} C_7\right)_{\alpha\beta}\;.\eqno(59)$$
Writing the BW equation
on $S_5$ in the form
$$\left( -\frac{i}{2} g_{ij} L_{ij} + \xi\right)_{\alpha\beta} \psi_{\beta\gamma} = 0,\eqno(60)$$
it follows from eqs. (59) and (60) (using eqs. (A.26) and (A.27) that
$$\xi \sigma = 0 \eqno(61)$$
$$L_{ij} \phi_{ij} = 0\eqno(62)$$
$$L_{ij} \tau_{ijk} = 0\eqno(63)$$
$$\epsilon_{ijk\ell mn} L_{jk} \tau_{\ell mn} = 0\eqno(64)$$
$$L_{ij} \sigma + \frac{i}{2} \epsilon_{ijk \ell mn} L_{k\ell} \phi_{mn} = 0\eqno(65)$$
$$i\left(L_{ki} \phi_{kj} - L_{kj}\phi_{ki}\right) + \xi\phi_{ij} = 0\eqno(66)$$
$$i\left(L_{\ell i}\tau_{\ell jk} +  L_{\ell j}\tau_{\ell ki} + L_{\ell k}\tau_{\ell ij}\right) + \xi\tau_{ijk} = 0.\eqno(67)$$
If $\xi \neq 0$, then by eq. (61), $\sigma = 0$ in which case $\phi_{ij}$ and $\tau_{ijk}$ decouple.  The equations for $\phi_{ij}$ are of exactly the same form as eqs. (50-51) on $S_4$ and hence can be treated in exactly the same way with gauge invariance occuring if $\xi = 4$. Eqs. (64) and (67) for the field $\tau_{ijk}$ possess the gauge invariance
$$\tau_{ijk} \rightarrow \tau_{ijk} + L_{ij}\rho_k + L_{jk} \rho_i + L_{ki} \rho_j\eqno(68)$$
provided again $\xi = 4$ and $\rho_i$ satisfies the equation
$$L_{\ell i}\rho_\ell - 2i\rho_i = 0.\eqno(69)$$
(This equation is akin to the gauge fixing condition used in ref. [3].) It does not appear to be feasable to devise a non-Abelian generalization of eq. (68).

\subsection{$S_6$}

On $S_6$, by eqs. (A.21) and (A.23), the only contributions to the BW wave function are
$$\psi_{\alpha\beta} = \sigma (C_7)_{\alpha\beta} + \tau_{ijk}\left(g_{ijk} C_7\right)_{\alpha\beta}\eqno(70)$$
so that using (A.25) in conjunction with the $S_6$ analogue of eq. (60)
$$\xi\sigma = 0\eqno(71)$$
$$L_{ij} \tau_{ijk} = 0\eqno(72)$$
$$L_{ij}\sigma + \frac{i}{2} \epsilon_{ijk\ell mnp} L_{kj} \tau_{mnp} = 0\eqno(73)$$
$$i\left(L_{\ell i} \tau_{\ell jk} + L_{\ell j} \tau_{\ell ki} + L_{\ell k} \tau_{\ell ij}\right) + \xi \tau_{ijk} = 0.\eqno(74)$$
These equations can be analyzed in much the same way as eqs. (63, 64, 67) for $\tau_{ijk}$ on $S_5$ were treated in the previous subsection.

\subsection{$S_8$}
We will now consider the BW equation on $S_8$ (The $S_7$ case is quite similar.).  On $S_8$, eq. (A.30 - A.32) imply that the wave function $\psi_{\alpha\beta}$ has the form
$$\psi_{\alpha\beta} = \sigma(C_9)_{\alpha\beta} + V_i \left(\Gamma_i C_9\right)_{\alpha\beta} + \Omega_{ijk\ell} \left(\Gamma_{ijk\ell}C_9\right)_{\alpha\beta}.\eqno(75)$$
Eqs. (A.33, A.34) now can be used to show that the BW equation of the form eq. (60) on $S_8$ leads to
$$\xi\sigma = 0\eqno(76)$$
$$i L_{ij} V_i + \xi V_j = 0\eqno(77)$$
$$L_{ij}\sigma - 12 L_{k\ell} \Omega_{ijk\ell} = 0\eqno(78)$$
$$\left(L_{ij} V_k + L_{jk} V_i + L_{ki} V_j\right) + \frac{1}{2} L_{mn} \Omega_{pqrs} \epsilon_{mnpqrsijk} = 0\eqno(79)$$
$$i\left(L_{mi} \Omega_{mjk\ell} + L_{mj} \Omega_{mk\ell i} + L_{mk} \Omega_{m\ell ij} + L_{m\ell}\Omega_{mijk}\right) + \xi \Omega_{ijk\ell} = 0\;.\eqno(80)$$

It is apparent that even when eq. (76) is satisfied by taking $\sigma = 0$, the fields $V_i$ and $\Omega_{ijk\ell}$ are coupled, making gauge invariance problematic.

\subsection{$S_{10}$}

Finally, we consider a model on $S_{10}$. On account of eq. (A.37), the BW wave function on $S_{10}$ has the form
$$\psi_{\alpha\beta} = V_i\left(G_iC_{11}\right)_{\alpha\beta} + \phi_{ij} \left(G_{ij}C_{11}\right)_{\alpha\beta}
+ \Lambda_{ijk\ell m} \left(G_{ijk\ell m}C_{11}\right)_{\alpha\beta} .\eqno(81)$$
By using eqs. (A.38 - A.40) in conjunction with the Bargmann-Wigner equation for $\psi_{\alpha\beta}$ in eq. (81), we find that
$$L_{ij} \phi_{ij} = 0\eqno(82)$$
$$iL_{ij} V_i + \xi V_j = 0\eqno(83)$$
$$i\left(L_{ki}\phi_{kj} - L_{kj}\phi_{ki}\right) +   \xi \phi_{ij} = 0\eqno(84)$$
$$\frac{1}{3} \left(L_{ij}V_k +  L_{jk}V_i + L_{ki} V_j\right) - 20 L_{mn} \Lambda_{mnijk} = 0 \eqno(85)$$
$$\left(L_{ij}\phi_{k\ell} - L_{kj}\phi_{i\ell} + L_{ki}\phi_{j\ell}+ L_{k\ell} \phi_{ij} - L_{i\ell} \phi_{kj}
+ L_{j\ell}\phi_{ki}\right)+ \frac{i}{4} L_{mn} \Lambda_{pqrs}\epsilon_{mnpqrsijk\ell} = 0 \eqno(86)$$
$$i\left(L_{pi} \Lambda_{pjk\ell m} + L_{pj} \Lambda_{pk\ell mi} + L_{pk} \Lambda_{p\ell mij}
+ L_{p\ell} \Lambda_{pmijk} + L_{pm} \Lambda_{pijk\ell}\right) + \xi\Lambda_{ijk\ell m} = 0.\eqno(87)$$

The fields $V_i, \phi_{ij}$ and $\Lambda_{ijk\ell m}$ are coupled in eqs. (82-87) in a way that appears to be inconsistent with gauge invariance.

\section{Quantization of a model on $S_2$}

We consider now the quantization of the action of eq. (23). We supplement this with the gauge fixing condition
$$L_{gf} = \alpha \phi^a_i L_i L_j \phi_j^a\eqno(88)$$
which is in part motivated by eq. (16).
Associated with this, a Faddeev-Popov ghost field must also be introduced as the gauge transformation of eq. (37) is non-Abelian,
$${\mathcal L}_{FP} = \overline{c}^a \left(\vec{L}^2 \delta^{ab} - i\epsilon^{apb} \vec{L} \cdot \vec{\phi}^p\right) c^b\; .\eqno(89)$$

The identity
$$\left(i\epsilon_{imj} L_m + \delta_{ij} + \alpha L_iL_j\right) \left(\frac{i}{\vec{L}^2}\epsilon_{jnk} L_n + \frac{1+\alpha}{\alpha(\vec{L})^2} L_jL_k\right) = \delta_{ik}\eqno(90)$$
permits one to derive a propagator for the field $\phi_i^a$.

Rather than using Feynman diagrams to compute radiative corrections to the model of eq. (33), we use the approach of ref. [8]. This entails splitting $\vec{\phi}^a$ into the sum of a background piece (also called $\vec{\phi}^a$) and a quantum piece; we are led to the one-loop effective action
$${\mathcal L}^{(1)} = \ln \left\{ det^{-1/2} \left[\left(i\epsilon_{imj}L_m + \delta_{ij} + \alpha L_iL_j\right) \delta^{ab}
+ \epsilon^{apb} \epsilon_{imj} \phi_m^p\right]
det \left(\vec{L}^2 \delta^{ab} - i\epsilon^{apb} \vec{L} \cdot \vec{\phi}^p\right)\right\}\; .\eqno(91)$$
In order to recast eq. (91) into a form in which calculations are feasable, we add on $\ln det^{-1/2}(i\epsilon_{inj}L_n)$\footnote{This piece is not gauge invariant, so employing operator regularization [8] is not appropriate.}. Since
$$i\epsilon_{imj} L_m \left[i\epsilon_{jnk}L_n + \delta_{jk} + \alpha L_jL_k\right]
= \delta_{ik} \vec{L}^2 - \left(L_kL_i + \alpha L_iL_k\right) - i\epsilon_{ik\ell} L_\ell\eqno(92)$$
we see that in the gauge $\alpha = -1$ this converts ${\mathcal L}^{(1)}$ in (91) to 
$${\mathcal L}^{(1)} = \ln \left\{ det^{-1/2} \left[\vec{L}^2 \delta_{ij} \delta^{ab} + i\epsilon^{apb} \left(L_j\phi_i^p - \delta_{ij} \vec{L} \cdot \vec{\phi}^p\right)\right]
det\left[\vec{L}^2 \delta^{ab} - i\epsilon^{apb} \vec{L} \cdot \vec{\phi}^p\right]\right\}\;.\eqno(93)$$
Using ``proper time"' [9] to express ${\mathcal L}^{(1)}$, we have
$${\mathcal L}^{(1)} = \int^\infty_0 \frac{dt}{t} tr \left\{ \frac{1}{2} e^{-H_\phi} - e^{-H_{gh}t}\right\}\eqno(94)$$
where $H_\phi$ and $H_{gh}$ are the arguments of the two determinants in eq. (94) respectively. To obtain the contribution of the two-point function to ${\mathcal L}^{(1)}$, we expand eq. (94) to second order in $\vec{\phi}^a$ using the Schwinger expansion [8-9]
$$tr e^{-(H_0+H_1)t} = tr\left\{ e^{-H_0t} + \frac{(-t)}{1} e^{-H_0t}H_1\right.\eqno(95)$$
$$\left. + \frac{(-t)^2}{2} \int_0^1 du\, e^{-(1-u)H_0t}H_1 e^{-uH_0t}H_1 + \ldots\right\}\; .\nonumber$$
Taking $H_0$ to be $\vec{L}^2$ in both $H_\phi$ and $H_{gh}$, eq. (94) gives rise to a contribution to ${\mathcal L}^{(1)}$ that is bilinear in $\vec{\phi}^a$
$${\mathcal L}_{\phi\phi}^{(2)} = \int_0^\infty \frac{dt}{t} \; \frac{(-t)^2}{2} \int_0^1 du\; e^{-(1-u)\vec{L}^2t} tr \left\{ \left( \frac{1}{2}\right)\left( i\epsilon^{apb}\left(L_j\phi_i^p\right.\right.\right.\eqno(96)$$
$$\left.\left.-\delta_{ij}\vec{L} \cdot \vec{\phi}^p\right)\right)e^{-u\vec{L}^2t}\left(i\epsilon^{bqa}\left(L_i\phi_j^q - \delta_{ji}\vec{L}\cdot \vec{\phi}^a\right)\right)\nonumber$$
$$\left. -\left(-i\epsilon^{apb}\vec{L}\cdot\vec{\phi}^p\right)
e^{-u\vec{L}^2t}
\left(-i\epsilon^{bqa}\vec{L}\cdot \vec{\phi}^q\right)
\right\}\nonumber$$
which simplifies to
$$ = \int_0^\infty dt\;t \int_0^1 du\; e^{-(1-u)\vec{L}^2t} tr\left\{ \frac{1}{2}\left[\left(L_i\phi_i^p\right)e^{-(1-u)\vec{L}^2t}\right.\right.\eqno(97)$$
$$\left. \left(L_j\phi_j^p\right) + \left(\phi_i^p L_i\right)e^{-(1-u)\vec{L}t}\left( \phi_j^pL_j\right)\right] \nonumber$$
$$\left. -\left[\left(L_i\phi_i^p\right)e^{-(1-u)\vec{L}^2t}\left(L_j\phi_j^p\right)\right]
\right\}\;.\nonumber$$

It is immediately evident that if the external field $\phi_i^a$ has vanishing angular momentum (so that $\left[L_i,\phi_j^p\right] = 0$) then eq. (97) gives ${\mathcal L}_{\phi\phi}^{(1)} = 0$ so that the two-point function vanishes. In general though, the functional traces in eq. (97) can be computed using a basis of angular momentum eigenstates [8,10]. We have, for example
$$tr\left(e^{-(1-u)\vec{L}^2t} L_i\phi_i^p e^{-u\vec{L}^2t} L_j\phi_j^p\right)
= \Sigma_{\ell = 0}^\infty \, \Sigma_{m = -\ell}^\ell\,\Sigma_{\ell^\prime = 0}^\infty\,\Sigma_{m^\prime = -\ell}^\ell \;e^{-[(1-u)\ell(\ell + 1)t+u\ell^\prime(\ell^\prime + 1)]t}
\eqno(98)$$
$$<\ell , m\left|L_i\phi_i^p\right|\ell^\prime ,m^\prime><\ell^\prime ,m^\prime\left|L_j\phi_j^p\right|\ell , m>.\nonumber$$
Using the conventions of ref. [11],
$$<\ell , m\left|L_i\phi_i^p\right|
\ell^\prime ,m^\prime> = 
<\ell  ,m\left|
\left(\frac{1}{2}\left(L_+\phi_-^p + L_-\phi_+^p\right) + L_3 \phi_3^p\right)
\right|\ell^\prime , m^\prime>\nonumber$$
$$=\frac{1}{2} \left[\sqrt{\ell(\ell + 1) - m(m - 1)} <\ell , m - 1\left|\phi_-^p\right|\ell^\prime ,m^\prime >\right. \eqno(99)$$
$$\left. + \sqrt{\ell(\ell + 1) - m(m + 1)} <\ell , m + 1\left|\phi_+^p\right|\ell^\prime ,m^\prime > + m <\ell , m\left|\phi_3^p\right|\ell , m >\right].\nonumber$$
If now
$$<\ell, m\left|\phi_i^p\right|\ell^\prime ,m^\prime > = \int d\eta <\ell , m\left|\eta><\eta\left|\phi_i^p\right|\eta><\eta\right|\ell^\prime ,m^\prime>\eqno(100)$$
with
$$<\ell, m\left|\eta> = Y_{\ell , m}(\eta)\right.\eqno(101)$$
being a spherical harmonic and
$$<\eta\left|\phi_i^p\right| \eta> =
{\displaystyle{\Sigma_{k=0}^\infty \Sigma_{n=-k}^k}} f_{k,n}^{p,i} Y_{k,n} (\eta)\eqno(102)$$
then eq. (100) reduces to
$$= \int d\eta {\displaystyle{\Sigma_{k,n}}} f_{k,n}^{p,i} Y_{\ell , m}(\eta)Y_{k,n}(\eta)\left[(-1)^{m^\prime} Y_{\ell^\prime , -m^\prime}(\eta)\right]\eqno(103)$$
which, using a standard formula [11], reduces to
$$= {\displaystyle{\Sigma_{k,n}}} f_{k,n}^{p,i} (-1)^m \left[\frac{(2\ell +1)(2\ell^\prime + 1)(2k+1)}{4\pi}\right]^{1/2} \left(\begin{array}{ccc}
\ell & \ell^\prime & k\\
0 & 0 & 0\end{array}\right)\left(\begin{array}{ccc}
\ell & \ell^\prime & k\\
m & -m^\prime & n\end{array}\right).\eqno(104)$$
Together, eqs. (98), (99) and (104) reduce eq. (97) to
$${\mathcal L}_{\phi\phi}^{(2)} = \frac{1}{2} \int_0^\infty dt\, t \int_0^1 du\,{\displaystyle{\Sigma_{\ell, m}\Sigma_{\ell^\prime, m^\prime}\Sigma_{k,n}}} e^{-[(1-u)\ell(\ell + 1) + u\ell^\prime (\ell^\prime + 1)]t}\nonumber$$
$$\left(\frac{(2\ell + 1)(2\ell^\prime + 1)(2k + 1)}{4\pi}\right)\left(\begin{array}{ccc}
\ell & \ell^\prime & k\\
0 & 0 & 0\end{array}\right)^2\nonumber$$
$$\left\{ - \left[ \frac{1}{2} \sqrt{\ell(\ell + 1) - m (m-1)} f_{k,n}^{p,-} (-1)^{m^\prime} \left(\begin{array}{ccc}
\ell & \ell^\prime & k\\
m-1 & -m^\prime & n\end{array}\right)\right.\right.\nonumber$$
$$+ \frac{1}{2} \sqrt{\ell(\ell + 1) - m (m+1)} f_{k,n}^{p,+} (-1)^{m^\prime} \left(\begin{array}{ccc}
\ell & \ell^\prime & k\\
m+1 & -m^\prime & n\end{array}\right)\nonumber$$
$$\left. +m \;f_{k,n}^{p,3} (-1)^{m^\prime} \left(\begin{array}{ccc}
\ell & \ell^\prime & k\\
m & -m^\prime & n\end{array}\right)\right]\nonumber$$
$$\left[ \frac{1}{2} \sqrt{\ell^\prime(\ell^\prime + 1) - m^\prime(m^\prime -1)} f_{k,n}^{p,-} (-1)^{m} \left(\begin{array}{ccc}
\ell^\prime & \ell & k\\
m^\prime-1 & -m & n\end{array}\right)\right.\nonumber$$
$$+ \frac{1}{2} \sqrt{\ell^\prime(\ell^\prime + 1) - m^\prime (m^\prime+1)} f_{k,n}^{p,+} (-1)^{m} \left(\begin{array}{ccc}
\ell^\prime & \ell & k\\
m^\prime +1 & -m & n\end{array}\right)\nonumber$$
$$\left. +  m^\prime f_{k,n}^{p,3} (-1)^{m} \left(\begin{array}{ccc}
\ell^\prime & \ell & k\\
m^\prime & -m & n\end{array}\right)\right]\nonumber$$
$$+ \left[ \frac{1}{2} \sqrt{\ell^\prime(\ell^\prime + 1) - m^\prime (m^\prime -1)} f_{k,n}^{p,+} (-1)^{m^\prime - 1} \left(\begin{array}{ccc}
\ell & \ell^\prime & k\\
m & -m^\prime + 1 & n\end{array}\right)\right. \nonumber$$
$$+ \frac{1}{2} \sqrt{\ell^\prime(\ell^\prime + 1) - m^\prime (m^\prime +1)} f_{k,n}^{p,-} (-1)^{m^\prime + 1} \left(\begin{array}{ccc}
\ell & \ell^\prime& k\\
m & -m^\prime - 1 & n\end{array}\right) \nonumber$$
$$\left. +  m^\prime f_{k,n}^{p,3} (-1)^{m^\prime} \left(\begin{array}{ccc}
\ell & \ell^\prime & k\\
m & -m^\prime & n\end{array}\right)\right] \nonumber$$
$$\left[ \frac{1}{2} \sqrt{\ell(\ell + 1) - m(m -1)} f_{k,n}^{p,+} (-1)^{m - 1} \left(\begin{array}{ccc}
\ell^\prime & \ell &  k\\
m^\prime & -m + 1 & n\end{array}\right)\right. \nonumber$$
$$+ \frac{1}{2} \sqrt{\ell(\ell + 1) - m (m+1)} f_{k,n}^{p,-} (-1)^{m+ 1} \left(\begin{array}{ccc}
\ell^\prime & \ell^\prime & k\\
m^\prime & -m - 1 & n\end{array}\right) \nonumber$$
$$\left.\left. + m \; f_{k,n}^{p,3} (-1)^m \left(\begin{array}{ccc}
\ell^\prime & \ell & k\\
m^\prime & -m & n\end{array}\right)\right]\right\}\;. \eqno(105)$$
If one orients the axes so that $f_{k,n}^{p,\pm} = 0$, then the remaining contribution to eq. (105) proportional to $f_{k,n}^{p,3}$ clearly vanishes due to a cancellation between the two functional determinants occurring in eq. (93).  Thus in the Feynman-like gauge $\alpha = 1$ there are no radiative corrections at one-loop order to the two-point function for the model of eq. (23).

We can also consider quantization of the $S_2$ model of eq. (27). If we again use the gauge fixing of eq. (88), then the bilinear term in the effective Lagrangian is
$$L^{(2)} = \phi_i^a \left(L^2 \delta_{ij} + i\epsilon_{ipj} L_p + \delta_{ij} - (1-\alpha) L_i L_j\right)\phi_j^a .\eqno(106)$$
In this case, the propagator for the field $\phi_i^a$ can be deduced from the identity
$$\left(L^2 \delta_{ij} + i\epsilon_{ipj} L_p - (1-\alpha) L_iL_j\right) \left(\frac{1}{L^2} \delta_{jk} - i\epsilon_{jqk}L_q\right. \eqno(107)$$
$$\left. + \frac{1}{\alpha L^2} \left(-\alpha + \frac{1-\alpha}{L^2}\right)L_jL_k\right) = \delta_{ik} .\nonumber$$
Radiative corrections can now be computed, provided the Faddeev-Popov contribution of eq. (89) is included.

To illustrate how one could deal with models on $S_n (n > 2)$, let us consider the action of eq. (36) on $S_3$. We employ a gauge fixing Lagrangian
$$L_{gf} = \phi_{ip} \left[ L_{qj}L_{qk} \left(2 \delta_{ij} \delta_{k\ell} - \frac{1}{2} \delta_{jk}\delta_{i\ell}\right)\right]\phi_{\ell p}.\eqno(108)$$
Although this gauge fixing Lagrangian is not a perfect square, it is consistent with the condition that
$$L_{ip} \phi_{pj} = 0.\eqno(109)$$
It resembles the gauge fixing Lagrangian employed in ref. [3] for a model of a vector gauge field on $S_4$.

It is convenient to supplement $L$ in eq. (26) with an extra term $2\lambda^2$. If we do this, then it follows that
$$L + L_{gf} = \left(\phi_{ij}, \lambda\right) 
\left(
\begin{array}{cc}
-\frac{i}{4} \left(\delta_{j\ell}L_{ik}\ldots\right) + \delta_{ij,k\ell}
& -\frac{1}{2} \epsilon_{ijpq} L_{pq}\\
+ \frac{1}{2}\left(\delta_{jk} L_{pi}L_{pk}\ldots\right)\\
-\frac{1}{2} L_{ij}L_{k\ell}\\
\frac{1}{2} \epsilon_{k\ell pq}L_{pq} & 2\end{array}\right)
\left(
\begin{array}{c}
\phi_{k\ell}\\
\lambda\end{array}\right)\eqno(110)$$
where
$$\delta_{j\ell}L_{ik} \ldots \equiv \delta_{j\ell} L_{ik} - \delta_{i\ell} L_{jk} + \delta_{ik} L_{j\ell} - \delta_{jk}L_{i\ell}\eqno(111)$$
and
$$\delta_{j\ell}L_{pi}L_{pk} \ldots \equiv \delta_{j\ell} L_{pi} L_{pk} - \delta_{i\ell} L_{pj} L_{pk}
 + \delta_{ik} L_{pj} L_{p\ell} - \delta_{jk}L_{pi}L_{p\ell}\eqno(112)$$
and
$$\delta_{ij,k\ell} \equiv \frac{1}{2} \left(\delta_{ik}\delta_{j\ell} - \delta_{i\ell}\delta_{jk}\right).\eqno(113)$$ 
Using the identity
$$\frac{1}{4} \epsilon_{ijpq} \epsilon_{mnrs} L_{pq} L_{rs} = \delta_{ij,k\ell} L_{pq} L_{pq} - \left(\delta_{im} L_{pj} L_{pn} \ldots \right) + L_{ij} L_{mn} + i \left(\delta_{im} L_{jn} \ldots \right)\eqno(114)$$
it can be shown that the inverse of the matrix $M$ appearing in eq. (110) is
$$M^{-1} = \frac{1}{D} \left(
\begin{array}{cc}
\delta_{ij,mn} & \frac{1}{4}\epsilon_{ijpq}L_{pq}\\
-\frac{1}{4} \epsilon_{mnpq} L_{pq} & \frac{1}{2}\end{array}\right)\eqno(115)$$
where $D = \left(1 + L^2/2\right)^{-1}$.
Eq. (115) essentially provides the propagators for the gauge model on $S_4$ appearing in eq. (26), allowing for a perturbative computation of radiative effects. Presumably other models on $S_n (n > 3)$ can be treated in a similar fashion.

\section{Discussion}

We have formulated models involving antisymmetric tensor fields on spherical spaces $S_n$, some of which have a non-Abelian generalization. Many open questions deserve attention. We have taken the radius $R$ of our spherical space to be 1; the locally flat limit $R \rightarrow \infty$ should be investigated. It is possible to investigage $BW$ wave functions with more than two spinor indices. Extensions to other spaces of constant curvature (such as (anti-) de Sitter space) would be of interest. The general structure of radiative corrections (renormalization, BRST identities etc.) merit consideration. If $\phi_{ij}^a$ is a gauge field, then it is possible to couple it to a matter field by replacing $L_{ij}$ by $L_{ij}\delta^{ab} + \epsilon^{apb}\phi_{ij}^b$ in the Lagrangian for the matter field. Supersymetric extensions may be possible. It is not clear if these models have phenomenological implications.

\section{Acknowledgement}

NSERC provided financial support. D.G.C. McKeon would like to thank NUI Galway and Ochanomizu University for their hospitality while much of this work was being done. R. and D. MacKenzie had helpful advice.

\section{Appendix}

The usual Pauli matrices $\tau_i$ satisfy
$$\tau_i\tau_j = \delta_{ij} + i\epsilon_{ijk}\tau_k .\eqno(A.1)$$
Dirac matrices in four and five dimensions are given by
$$\gamma_{1,2,3} = \left(\begin{array}{cc}
0 & i\tau_{1,2,3} \\
-i \tau_{1,2,3} & 0\end{array}\right)\;\;\; 
\gamma_4 = \left(\begin{array}{cc}
0 & 1 \\
1 & 0\end{array}\right)\;\;\;
\gamma_5 = \left(\begin{array}{cc}
-1 & 0 \\
0 & 1\end{array}\right) = \gamma_1\gamma_2\gamma_3\gamma_4 .\eqno(A.2)$$
Spin matrices in four and five dimensions are given by
$$\Sigma_{ab} = -\frac{i}{4}\left[\gamma_a,\gamma_b\right] \equiv -\frac{i}{2} \gamma_{ab};\eqno(A.3)$$
these satisfy
$$\left[\Sigma_{ab}, \Sigma_{cd}\right] = i\left(\delta_{ac}\Sigma_{bd} - \delta_{ad}\Sigma_{bc} + \delta_{bd}\Sigma_{ac} - \delta_{bc}\Sigma_{ad}\right).
\eqno(A.4)$$
The generators
$$L_{ab} = -i\left(\eta_a\frac{\partial}{\partial\eta_{b}} - \eta_b \frac{\partial}{\partial\eta_{a}}\right)\eqno(A.5)$$
also satisfy the algebra of eq. (A.4).

A useful identity in five dimensions is
$$\left\{\Sigma_{ab}, \Sigma_{cd}\right\} = \frac{1}{2} \left(\delta_{ac}\delta_{bd} - \delta_{ad}\delta_{bc}\right) - \frac{1}{2} \epsilon_{abcde}\gamma_e\eqno(A.6)$$
while in four dimensions,
$$\left\{\Sigma_{ab}, \Sigma_{cd}\right\} = \frac{1}{2} \left(\delta_{ac}\delta_{bd} - \delta_{ad}\delta_{bc}\right) - \frac{1}{2} \epsilon_{abcd}\gamma_5\eqno(A.7)$$
$$\left[ \Sigma_{ab}, \gamma_c \right] = i \left(\delta_{ac}\gamma_b - \delta_{bc}\gamma_a\right)\eqno(A.8)$$
$$\left\{ \Sigma_{ab}, \gamma_c \right\} = i \epsilon_{abcd}\gamma_d \gamma_5\eqno(A.9)$$
and
$$\gamma_a\gamma_b\gamma_c = \delta_{ab}\gamma_c - \delta_{ac} \gamma_b + \delta_{bc}\gamma_a - \epsilon_{abcd}\gamma_\alpha\gamma_5.\eqno(A.10)$$

In three dimensions if
$$C_3 = \tau_2 = -C_3^T ,\eqno(A.11)$$
then
$$\left(\tau_i C_3\right)^T = \left(\tau_iC_3\right).\eqno(A.12)$$
In four and five dimensions we take
$$C_5 = \gamma_1\gamma_3 = -C^T =-C^{-1}\eqno(A.13)$$
so that
$$\left(\gamma_aC_5\right)^7 = -\left(\gamma_aC_5\right).\eqno(A.14)$$
The only symmetric matrices in four dimensions are
$$\left(\gamma_a\gamma_5C_5\right)^T = \left(\gamma_a\gamma_5C_5\right)\eqno(A.15)$$
and
$$\left(\Sigma_{ab} C_4\right)^T = \left(\Sigma_{ab}C_5\right)\eqno(A.16)$$
while in five dimensions the only symmetric matrix is
$$\left(\Sigma_{ab}C_5\right)^T = \left(\Sigma_{ab}C_5\right).\eqno(A.17)$$

The Dirac matrices in six and seven dimensions are given by
$$g_{1,2, \ldots , 5} = \left(\begin{array}{cc}
0 & i\gamma_{1,2, \ldots , 5}\\
i\gamma_{1,2, \ldots , 5} & 0\end{array}\right) \;\;\;
g_6 = \left(\begin{array}{cc}
0 & 1\\
1 & 0\end{array}\right)\;\;\;
g_7 = \left(\begin{array}{cc}
-1 & 0\\
0 & 1\end{array}\right) = ig_1 \ldots g_6\;.\eqno(A.18)$$

The independent $8 \times 8$ Dirac matrices in seven dimensions are $1$, $g_a$, $g_{a,b}$ and $g_{abc}$ where
$$g_{ab} = \frac{1}{2!}\left(g_ag_b -g_bg_a\right)\eqno(A.19)$$
and
$$g_{abc} = \frac{1}{3!}\left(g_ag_bg_c +\;{\rm{antisymmetric\;terms}}\right).\eqno(A.20)$$

If
$$C_7 = g_2g_4g_5 = C_7^T = -C_7^{-1}\eqno(A.21)$$
then in six and seven dimensions
$$\left(g_a C_7\right)^T = -\left(g_aC_7\right).\eqno(A.22)$$
The symmetric $8 \times 8$ matrices in seven dimensions are $C_7$ and $g_{abc}$;
$$\left(g_{abc}C_7\right)^T = \left(g_{abc}C_7\right).\eqno(A.23)$$
This implies that in six dimensions, the symmetric $8 \times 8$ matrices are $C_7$, $g_{abc}$ and $g_{ab}g_7$;
$$\left(g_{ab}g_7C_7\right)^T = \left(g_{ab}g_7C_7\right).\eqno(A.24)$$

A useful identity in seven dimensions is
$$g_{mn}g_{pqr} = -\left(\delta_{mp}\delta_{nq}g_r + {\rm{antisymmetric\;in\;}}(mn)(pqr)\right)\eqno(A.25)$$
$$- \left(\delta_{mp}g_{nqr} + {\rm{antisymmetric\;in\;}}(mn)(pqr)\right)\nonumber$$
$$+ \frac{i}{2} \epsilon_{mnpqrab}g_{ab}.\nonumber$$

In six dimensions, we have occasion to use
$$g_{mn}g_{ab}g_7 = -\left(\delta_{ma}\delta_{nb} - \delta_{mb}\delta_{na}\right)g_7\eqno(A.26)$$
$$-\left(\delta_{ma}g_{nb} - \delta_{mb}g_{na} + \delta_{nb}g_{ma} - \delta_{na}g_{mb}\right)g_7\nonumber$$
$$+ \frac{i}{2} \epsilon_{mnabpq}g_{pq}\nonumber$$
and
$$g_{mn}g_{pqr} = -\left(\delta_{mp}\delta_{nq} g_r+  + {\rm{antisymmetric \;in}} (mn)(pqr)\right)\nonumber$$
$$-\left(\delta_{mp}g_{nqr} + {\rm{antisymmetric\; in}} (mn)(pqr)\right)\nonumber$$
$$+ i \epsilon_{mnpqa}g_ag_7.\eqno(A.27)$$

In eight and nine dimensions, we have the $16 \times 16$ matrices
$$\Gamma_a = \left(\begin{array}{cc}
0 & ig_a\\
-ig_a & 0\end{array}\right)\;\;
(a = 1 \ldots 7),\;\;
\Gamma_8 = \left(\begin{array}{cc}
0 & 1\\
1 & 0\end{array}\right),\;\;
\Gamma_9 = \left(\begin{array}{cc}
-1 & 0\\
0 & 1\end{array}\right) = -\Gamma_1 \ldots \Gamma_8\;.\eqno(A.28)$$
The independent $16 \times 16$ matrices in nine dimensions are $1$, $\Gamma_a$, $\Gamma_{ab}$, $\Gamma_{abc}$ and $\Gamma_{abcd}$ where for $n = 2,3,4$
$$\Gamma_{a_{1}\ldots a_{n}} = \frac{1}{n!}\left(\Gamma_{a_{1}}\Gamma_{a_{2}}\ldots \Gamma_{a_{n}} + {\rm{antisymmetric\;terms}}\right).\eqno(A.29)$$
If now
$$C_9 = \Gamma_1\Gamma_3\Gamma_6\Gamma_7 = C_9^T = C_9^{-1}\eqno(A.30)$$
then the symmetric matrices in nine dimensions are $C_9$ and
$$\left(\Gamma_aC\right)^T = \Gamma_aC\eqno(A.31)$$
$$\left(\Gamma_{abcd}C\right)^T = \Gamma_{abcd}C.\eqno(A.32)$$
Two useful identities in nine dimensions are
$$\Gamma_{mn}\Gamma_a = -\left(\delta_{ma} \Gamma_n - \delta_{na}\Gamma_b\right) + \Gamma_{mna}\eqno(A.33)$$
and
$$\Gamma_{mn}\Gamma_{abc} = -\left(\delta_{ma}\Gamma_{nbcd} + 
{\rm{antisymmetric\; in}}\; (mn)(abcd)\right)\nonumber$$
$$-\left(\delta_{ma}\delta_{nb}\Gamma_{cd} + {\rm{antisymmetric\; in}}\; (mn)(abcd)\right)\nonumber$$
$$+ \frac{1}{3!} \epsilon_{mnabcdpqr}\Gamma_{pqr}.\eqno(A.34)$$

The $32 \times 32$ Dirac matrices in ten and eleven dimensions
are
$$G_a = \left(\begin{array}{cc}
0 & i\Gamma_a\\
-i\Gamma_a & 0\end{array}\right)\;
(a = 1 \ldots 9),\;\;
G_{10} = \left(\begin{array}{cc}
0 & 1\\
1 & 0\end{array}\right),\;\;{\rm{and}}\;\;
G_{11} = \left(\begin{array}{cc}
-1 & 0\\
0 & 1\end{array}\right) = -iG_1 \ldots G_{10};\eqno(A.35)$$
a complete set of $32 \times 32$ matrices are $1$, $G_a$, $G_{ab}$, $G_{abc}$, $G_{abcd}$ and $G_{abcde}$ where $G_{a{_1}},\ldots a_n$ is defined in analogy with $\Gamma_{a{_1}} \ldots a_n$ in eq. (A.29).

If now
$$C_{11} = G_2G_4G_5G_8G_9 = -C_{11}^T = C_{11}^{-1}\eqno(A.36)$$
then it can be shown that
$$\left( G_{a_1} \ldots a_nC_n\right)^T = (-1)^{\frac{n^2+n+2}{2}}
\left( G_{a{_1}} \ldots a_nC_{11}\right)\eqno(A.37)$$
so that symmetric matrices occur if $n = 1,2,5$.

Useful eleven dimensional identities are
$$G_{mn} G_a = -\left(\delta_{ma} G_n - G_{na}G_m\right) + G_{mna}\eqno(A.38)$$
$$G_{mn} G_{ab} = -\left(\delta_{ma} \delta_{nb} - \delta_{mb}\delta_{na}\right) - \left( \delta_{ma} G_{nb} - \delta_{na}G_{mb}\right.\nonumber$$
$$\left. + \delta_{nb} G_{ma} - \delta_{mb}G_{na}\right) + G_{mnab}\eqno(A.39)$$
$$G_{mn} G_{abcde} = -\left(\delta_{ma}\delta_{nb} G_{cde} +
{\rm{antisymmetric\; in}} \;(mn)(abcde)\right)\nonumber$$
$$-\left(\delta_{ma}G_{nbcde} +  {\rm{antisymmetric\; in}}\; (mn)(abcde)\right)\nonumber$$
$$+ \frac{i}{4} \epsilon_{mnabcdepqrs} G_{pqrs}.\eqno(A.40)$$

\end{document}